 \documentstyle[preprint,aps,psfig]{revtex}  
\tightenlines

\begin{document}
\draft
\title{Backbending in Dy isotopes within the Projected Shell Model} 

\author{
V. Vel\'azquez$^1$, J. G. Hirsch$^2$, Y. Sun$^3$ and M. W. Guidry$^3$\\
{\small\it $^1$ Departamento de F\'{\i}sica, Centro de Investigaci\'on y
Estudios Avanzados del IPN,} \\
{\small\it A.P. 14-740, 07000 M\'exico D.F., M\'exico    }  \\
{\small\it $^2$ Instituto de Ciencias Nucleares, UNAM,} \\
{\small\it Circuito Exterior C.U.,A.P. 70-543, 04510 M\'exico D.F.,
M\'exico    }  \\
{\small\it $^3$Department of Physics and Astronomy, University of Tennessee,} \\
{\small\it Knoxville, Tennessee  37996, USA} \\
} 

\maketitle

\begin{abstract}
A systematic study of the yrast band in $^{154-164}Dy$ isotopes using the
Projected Shell Model is presented. 
It is shown that, in the context of the present model,
enlarging the mean field deformation by about $20\%$ allows a very good
description of the spectrum of yrast band in these isotopes. 
The dependence of the $B(E2)$ values on angular momentum is also better
described when larger deformations are used.
The observed oscillation of g-factors at low spin states 
remains an open question for this model.  
\end{abstract}

\bigskip

\newpage
\section{Introduction}

Backbending in the moment of inertia \cite{bb1,bb2}, which is a common 
phenomenon for many heavy and deformed nuclei,  
is understood as a consequence of crossing 
between two rotational bands, one being the ground band and the other
having a pair of aligned high-$j$ intruder particles \cite{SS72}.
Its magnitude is related to the crossing angle between the
crossing bands \cite{Har91}. 
A large crossing angle implies that the
bands interact over a narrow angular momentum region, and a sharp backbending
usually takes place. 
A small crossing angle implies an interaction which
spreads its influence along a wide angular momentum region, therefore
producing a smooth backbending; 
in some cases, only an upbending instead of a backbending is seen.
The above physical picture of the band crossing can be clearly seen
in the Projected Shell Model (PSM) \cite{Har95}, which 
has been applied successfully
to the description of the energy spectra and electromagnetic transitions
in deformed nuclei. 

High spin properties along the yrast line in the Dy isotope chain were 
studied by various cranked mean-field theories, with and
without particle number projection \cite{Ces91}. 
It was concluded in Ref. \cite{Ces91} that, while the average behavior of
the experimental data can be reproduced by cranking
models, a microscopic interpretation of some 
observations, in particular those related to band crossings,
requires investigations going beyond the
mean field approximation.
More recent theoretical work \cite{Sun94} used the PSM 
to go beyond the mean field,
and the results were compared with the spectrum, the 
$B(E2)$ and the gyromagnetic factor 
(g-factor) data existing at that time. 
In Ref. \cite{Sun94}, an improvement in overall description 
of the Dy isotopes was found but, in
many cases, the PSM seemed to exaggerate the backbending. 

The main goal of
the present article is to demonstrate that, by increasing the mean field
deformation by 20\% on average, both the backbending and the angular 
momentum dependence of the quadrupole moments in the Dy chain can be
quantitatively described within the PSM.
It will be shown that the increase in 
deformation is reflected in a rearrangement of single
particle states around the Fermi level 
that smoothes backbending curves, in agreement
with the experimental observation.
Nevertheless, 
for lighter isotopes, the oscillation in the recent g-factor data 
around angular momentum $I = 6\hbar$ and the reduction 
in $B(E2)$ values at high angular
momenta cannot be reproduced by the PSM within the current model space.  

The paper is arranged as follows: 
In section II, we outline the PSM. 
Interested readers 
are kindly referred to
the review article \cite{Har95} 
and the corresponding computer codes \cite{Sun97} in the literature. 
In section III, we introduce the mean field where our shell model 
space is truncated and discuss how it is related to
our final results. 
Discussions on the spectra in terms of the
backbending plots, the $B(E2)$'s and the 
g-factors for the $^{154 - 164}$Dy isotopes are presented
in sections IV, V and VI, respectively,
where a large body of the experimental high spin data is  
compared systematically with the theory. Finally, a summary is given 
in section VII.

\section{The Model}

The PSM allows a many body quantum mechanical description of atomic
nuclei, while avoiding the use of an extremely large Hilbert spaces
commonly required in spherical shell model calculations \cite{Har95}. 
Assuming an axial symmetry in the single particle potential the PSM uses a
multi-quasiparticle (qp) basis $ |\phi_{\kappa}>$
built on the Nilsson + BCS mean field.
Its elegance and efficiency lies in the use of angular momentum projection
through the angular momentum projection operator
$\hat{P}^I_{KK'}$  to carry the multi-qp states from the intrinsic to
the laboratory system \cite{Rin80}.

The present type of PSM Hamiltonian is schematic, 
with three classes of residual
interactions: quadrupole-quadrupole,  monopole pairing and quadrupole
pairing. This is basically the quadrupole plus pairing Hamiltonian and has 
been widely used in nuclear structure studies \cite{Rin80}.
The Hamiltonian can be expressed as

\begin{equation}
\hat{H}= \hat{H_0}-\frac{\chi}{2} \sum_ {\mu} \hat{Q}^+_{\mu}
\hat{Q}_{\mu}-G_M \hat{P}^+ \hat{P} - G_{Q} \hat{P}^+_{\mu} \hat{P}_{\mu}, 
\label{ham}
\end{equation} 
where $\hat{H_0}$ is the spherical single particle Hamiltonian. 
The Hamiltonian (\ref{ham}) is diagonalized in the angular momentum projected
multi-qp
basis \{ $\hat{P}^I_{M K}|\phi_{\kappa}\rangle$\}. 
The eigenvalue equation is:

\begin{equation}
\sum _{\kappa ' K'} (H^I _{\kappa K \kappa 'K'}-EN^I
_{\kappa K \kappa
'K'})F^{I E}_{ \kappa 'K'} = 0, 
\label{eigen}
\end{equation}
with the normalization condition

\begin{equation}
\sum _{\kappa K \kappa 'K'} F^{I E}_{ \kappa K}N^I _{\kappa K
\kappa 'K'}
F^{I E'}_{ \kappa 'K'} = \delta_{E E'},
\label{orto}
\end{equation}
where

\begin{equation}
\begin{array}{ll}
H^I _{\kappa K \kappa 'K'} = &\langle \phi_{\kappa}|\hat{H}
\hat{P}^I_{ K
K'}|\phi_{\kappa'}\rangle\\
N^I _{\kappa K \kappa 'K'} = &\langle\phi_{\kappa}|\hat{P}^I_{ K
K'}|\phi_{\kappa'}\rangle.   
\label{norm}
\end{array}
\end{equation}
After solving equation (\ref{eigen}), we obtain 
the normalized eigenstates with energy $E$:

\begin{equation}
 |\Psi^E_{IM}\rangle = \sum _{\kappa K} F^{I E}_{ \kappa K}
\hat{P}^I_{MK}|\phi_{\kappa}\rangle .
\label{wf}
\end{equation}
Electromagnetic properties of the nuclei are then
computed by using the wave functions obtained above.

\section{The Mean field and the residual interaction strengths}

In the PSM, the multi-qp basis $|\phi_\kappa\rangle$ is built on 
qp excitations from the Nilsson + BCS vacuum.
The shell model basis (the angular momentum projected multi-qp
states) is then truncated according to energy excitation
from the qp vacuum. Because for most quantities of interest in
spectroscopy only states relatively near the Fermi surface are important, 
the usual dimension where diagonalization is carried out is 
about 100 in realistic calculations. 
Since this is a truncated theory, the quality of a calculation 
is related to the quality of the mean field, particularly 
the qp states
around the Fermi level if the interest in question is the yrast line. 
Given the  mean field is often used in describing heavy and deformed
nuclei considerable knowledge exists concerning the qp
structure near the Fermi surface \cite{Rin80}. 

In the PSM, the quadrupole-quadrupole strength 
in Eq. (\ref{ham}) is obtained
self-consistently from the mean field quadrupole deformation $\epsilon_2$,
while the pairing and quadrupole pairing strengths 
in Eq. (\ref{ham}) are taken from
systematics \cite{Har95}. In a previous study of $^{160}$Dy, it has been
shown that low energy rotational bands can only be
reproduced if these self-consistent values are used\cite{Vel98}. 

The monopole and quadrupole-pairing interaction strengths $G_M$ and $G_Q$ used
in this work are

\begin{eqnarray}
G_M = (20.12 \mp 13.13 \frac{N-Z}{A})A^{-1} \\
G_Q = 0.18 ~G_M, 
\end{eqnarray}
where the minus(plus) sign is applied to neutrons (protons).
These are the same values used in Refs. \cite{Har91,Sun94}. 

The deformation $\epsilon_2$ can be taken from tables \cite{Atomic95}
which are extracted from experimental 
$B(E2,2^+_1 \rightarrow 0^+_{gs})$ values by using a
geometrical model. 
These values are very similar to those calculated from 
mean field models \cite{Beng86,Lamm69}. This is expected because in 
mean field models the relation between the deformation and the
electric quadrupole moment is constructed so 
that the experimental $B(E2)$ should be reproduced. 
However, it should be kept in mind that deformation is a 
model-dependent concept, while the $B(E2)$ is directly related to 
measured quantities. 
 
We found that the use of an enlarged deformation $\epsilon_2'$ strongly
improves the description of the spectra and $B(E2)$ values within the PSM 
for the Dy chain. 
The same effect is operative in other rare earth isotope chains
\cite{Vel99}.
A detailed analysis of the changes in the
mean field energies and quasiparticle occupations introduced by this
modified deformation, as well as its effects in the backbending plot, are
presented in the following sections.

Table 1 shows the deformations used in this article for the different Dy
isotopes, which are listed in the first column. In the second and third
columns, we list the
deformations reported in the literature \cite{Atomic95} and the enlarged
ones used in our calculation. The latter ones were obtained
by looking for the best available description of the energy spectra and
backbending plots for each isotope. Once the deformation was fixed, the
quadrupole-quadrupole interaction strength was obtained selfconsistently
\cite{Har95,Vel98}. An overall deformation increase of around 20\% was
found necessary to best reproduce the experimental data od Dy isotopes.

\section{The Backbending plots}

The angular frequency $\omega$, defined as
$\hbar \omega(I) = {\frac 1 2} [E(I+2) - E(I)]$, is
plotted as a function of the spin $I$ for the yrast states in Fig. 1.
Results for $^{154,156,158,160,162,164}$Dy are presented in figures
(a,b,c,d,e,f) respectively. In each plot the full line corresponds to
standard deformation $\epsilon_2$  and the dotted line to the enlarged
deformation $\epsilon_2'$. Diamonds represent the
experimental data, taken from Ref. \cite{Emling84,Azgui85}.

It can be seen that in each case a larger
deformation is associated with a smoother curve, thus better reproducing the
experimental data. For the first four Dy isotopes, the PSM calculation with
standard deformation predicts a second backbending \cite{Sun94}, 
which is visualized as a second
minimum seen in Fig. 1 and  
has no experimental counterpart. 
This prediction disappears in the calculation with enlarged deformation. 
Only in $^{154}$Dy does the
experimental 
yrast band exhibit a very sharp 
second backbending at $I = 30 \hbar$, which is not
predicted by either the standard or the enlarged deformation calculations.

The same conclusion is emphasized when presented in a plot with twice
the moment of
inertia $2 \Theta$ vs. the square of the angular frequency 
(the backbending plots),
as it is shown in Fig. 2. In these plots the backbending 
features are emphasized and the
 improvement obtained in the description of the experimental
data using the enlarged deformation is clear. 
While for the heavier Dy isotopes the moment of inertia changes very
slowly as a function of the angular frequency, a strong stretching effect is
apparent for the lighter ones. This implies that, in the case of
$^{154,156}$Dy at small rotational
frequencies, the moment of inertia has drastic changes which would
require for its proper description the inclusion of 
a richer mean field basis, rather than a geometrically fixed deformation
with axial symmetry.
Due to the absence of this ingredient in the present calculation,
the moment of inertia at low angular momenta 
in these two isotopes is not well described.

In order to obtain a deeper understanding of the results seen in the
backbending plots with different deformations, in Fig. 3 we present
the unperturbed band diagram of $^{158}$Dy. In this figure the
band energy, defined as \cite{Har95}

\begin{equation}
E_\kappa^I = 
{\frac {H^I _{\kappa K \kappa K}} {N^I _{\kappa K \kappa K}}} =
{\frac { \langle\phi_{\kappa}|\hat{H} \hat{P}^I_{ K
K}|\phi_{\kappa}\rangle}
{ \langle\phi_{\kappa}|\hat{P}^I_{ K K}|\phi_{\kappa}\rangle}}, 
\end{equation}
is plotted as a function of angular momentum $I$. For the
standard deformation (left figure in Fig. 3) 
the 2-qp band (labeled by number 2) that is the first one to cross the
ground state
band corresponds to a configuration with particles from smaller
$K$-orbits ($K = 1/2$ and 3/2).
In the case of the
enlarged deformation (right figure in Fig. 3) 
the relevant 2-qp band (again labeled as 2) has particles from larger 
$K$-orbits ($K = 3/2$ and 5/2). 
In the first
case the crossing angle between the unperturbed ground and that 2-qp  
band at $I = 12 \hbar$, which we loosely define as the angle
between the tangents of both curves at the crossing point
\cite{Har91}, is large, and is
reflected in Fig. 2c as an exaggerated zig-zag effect in the full line.
On the other hand, a smaller crossing angle at $I=16\hbar$ for the
enlarged deformation case is behind the smooth behavior of the dotted
curve, which precisely reproduces the experimental information.

At the standard deformation there are three neutron single particle levels
that contribute to the two quasiparticle bands active in the backbending
region. They are the Nilsson states with important spherical components
N = 6, l=i, j=13/2, m= 5/2, -3/2, 1/2, denoted [6 ~i13/2 ~5/2]n, [6 ~i13/2
~-3/2]n and [6 ~i13/2 ~1/2]n respectively. When the deformation is
increased, 
only the first two levels are relevant for backbending.
In Fig. 4, some neutron Nilsson single particle levels 
close to the Fermi level (indicated in
each figure as a black dot) are plotted. The rearrangements of the
relative position of
these levels explains the displacement to higher energies of many two-qp
bands for the enlarged deformation, with the consequent changes in the
band crossings and the backbending plots. 

Notice that, for $\epsilon_2 \approx 0.3$ and N = 92, 
there is a large gap and a sparse number of single particle
states for 92 $\leq$ N$\leq$ 96. For this reason, the excitation
energy and backbending plots for $^{160,162,164}$Dy shown in Fig. 1 d,e,f
and Fig. 2 d,e,f are normally behaved, with no sudden changes in their
slopes.
    
Until now, we have argued the necessity of changing the single particle
distributions by shifting the Fermi level to that corresponding to a
larger deformation in the standard Nilsson diagram. 
It should be noted that the parameters used to generate the
Nilsson diagram in the present paper
were fitted nearly 30 years ago \cite{nkm}, when not many accurate and
systematic high-spin data were available. 
An alternative of 
improvement the single particle distribution 
is to modify the standard Nilsson parameters, i.e. changing the local single 
particle distribution in the standard Nilsson diagram. 
It has been noticed recently \cite{Sun98} 
that the standard Nilsson parameters for the
proton $N = 5$ shell need to be modified to reproduce 
the newest high spin data.
 
In an early PSM work \cite{Har91}, the modified
set of Nilsson parameters \cite{Ben90} for lighter Er and Yb isotopes was used, 
while the standard Nilsson parameters \cite{nkm}
were employed for heavier isotopes. Namely, if the authors of \cite{Har91} 
insisted on assuming the deformations commonly
found in the literature, they had to use 
different sets of the Nilsson parameters
for lighter and heavier isotopes. 
We have tested our prescription of increasing deformation 
in Er and Yb nuclei, and found that one is able to use one unified 
(standard) set of Nilsson parameters to span a deformed basis for the PSM, 
with the lighter isotopes requiring a larger deformation increase.

\section{Reduced quadrupole moment and B(E2) transitions}

The $B(E2; I_i \rightarrow I_f)$ transition probabilities from initial
state $(\sigma_i,I_i)$ to final state $(\sigma_f,I_f)$ are given by
\cite{Rin80}

\begin{eqnarray}
B(E2;I_i \rightarrow I_f)=\frac{2I'+1}{2I+1}
|\langle\Psi_{I'}||\hat{Q}_{2}||\Psi_I\rangle |^2, 
\end{eqnarray}
where \cite{Har95}

\begin{eqnarray}
\langle \Psi_{I'}||\hat{Q}_{2}||\Psi_I \rangle =
\sum_\nu\{\sum_{\kappa \kappa'}
(IK' -\nu',\lambda \nu| I'K') 
\langle \phi_{\kappa'}|\hat{Q}_{2\nu}\hat{P}^I_{K'- \nu
K}|\phi_{\kappa} \rangle F^{I'}_{\kappa'}F^{I}_{\kappa}\}.
\end{eqnarray}
Here $F^{I'}_{\kappa'}$ and $F^{I}_{\kappa}$ are, respectively, the
PSM eigenvectors for spins $I$ and $I'$, as calculated in Eq. (\ref{eigen}),
and the electric quadrupole operator $Q_{2\nu}$ is defined as \cite{Rin80}

\begin{eqnarray}
Q_{2\nu} = e_p Q^p_{2\nu} + e_n Q^n_{2\nu}~,\hspace{1cm}
Q^{p (n)}_{2\nu} = e \sum_{i=1}^{Z (N)} r_i^2 Y_{2\nu}(\theta_i,\phi_i).
\end{eqnarray}

The effective charges $e_p$ and $e_n$ for protons and neutrons were taken as
$e_p=1.5$ and $e_n=0.5$ in previous works \cite{Har95,Sun94}. We have used
the same values in the calculations involving the standard deformation
in this paper.
However, when using the enlarged deformation, it proved 
necessary to 
use a slightly smaller effectives charges, as prescribed for example in
Ref. \cite{Bhatt92}:
\begin{equation}
e_p=e(1+\frac{Z}{A}),~~~~
e_n=e(\frac{Z}{A})
\end{equation}
For $^{158}$Dy their numerical values are $e_p=1.41$ and $e_n=0.41$.

Fig. 5 exhibits the $B(E2; I_i \rightarrow I_f)$ values for the six Dy
isotopes under study. Again the full lines represent the results obtained
with the standard deformation and the dotted lines those obtained with the
enlarged one. Experimental data are presented with their error bars. 
There are strong reductions in the $B(E2)$ values predicted around
$I = 14\hbar$ using the standard deformation that clearly contradict the
experimental data. This contradiction is now removed 
in the calculation with the
enlarged deformation. 

Another quantity of interest is the reduced transition quadrupole moment.
It is defined as \cite{Har95,Emling84}

\begin{equation}
Q_t(I \rightarrow I')=\frac{1}{(I0,20|I'0)} 
\sqrt{
\frac{2I+1}{2I'+1}B(E2;II')}. 
\end{equation}
For a rigid rotor with axial symmetry, it is equivalent, up to a sign, to
the static quadrupole moment. In this special case \cite{Har95}
$$Q_t(I\rightarrow I-2)/Q_t(2\rightarrow 0) = 1~.$$ 
An analysis of this quotient shows the extent to which
the yrast band $B(E2)$ values correspond to a symmetric axial rotor.
Cescato {\it et al.} showed \cite{Ces91} that the cranking models
give a rigid rotor behavior for any spin and can not explain the
experimental fluctuations in the quadrupole moment 
as function of angular momentum. 

The Fig. 6(a,b,c,d,e,f) shows the behavior of the ratio
$Q_t(I\rightarrow I-2)/Q_t(2\rightarrow 0)$ as function of the total
angular momentum. 
As seen in the figure, 
the theoretical predictions exhibit a sudden change
in magnitude at those angular momenta where backbending takes place. 
As was mentioned in the discussion of $B(E2)$ values, at standard
deformations the PSM predicts more variations than observed, while with
the enlarged deformation the agreement with the experimental data is
improved. 

The origin of the improved agreement in the transition quadrupole matrix
elements is similar to that of the improved backbending behavior. A
smoothing of the crossing interaction will generally tend to suppress
sharp
drops of the transition matrix elements at the first backbending.

The reduction of the $B(E2)$ values observed in
$^{156,158}$Dy for $I > 20 \hbar$ and the corresponding reduction in
quadrupole moments can not be reproduced in any of
the present calculations which have included only a few states near the
Fermi surface.
Qualitatively, this would require a more extensive PSM description
including in the Hilbert space more 2- and 4-qp states,
and possibly even higher orders of qp states, to
account for this phenomenon of gradual loss of collectivity.

\section{The Gyromagnetic Factors}

Gyromagnetic factors are very sensitive to the particle alignment
processes, thus allowing us to disentangle the origin of the crossing band
at spin around $16-18 \hbar$.
The magnetic moment $\mu$ of a state $(\sigma, I)$ is defined by
\begin{equation}
  \mu(\sigma, I)  =  \sqrt{\frac{4\pi}{3}} <\sigma, II|{\cal M}_{10}|\sigma, II>
  = \frac{[4 \pi I]^{1/2}}{[3(I+1)(2I+1)]^{1/2}} <\sigma, I||{\cal
M}_{1}||\sigma, I>,
 \label{mag}
\end{equation}
the  operator  ${\cal M}_{10}$ is  given by
\begin{equation}
{\cal M}_{10}  =  \mu_N \sqrt{\frac{3}{4\pi}} \sum_{i=1}^{A}
  g^{(i)}_l l^{(i)}_z + g^{(i)}_s s^{(i)}_z
               =  \mu_N \sqrt{\frac{3}{4\pi}} \sum_{\tau =p,n}
   g^{\tau}_l L^{\tau}_z + g^{\tau}_s S^{\tau}_z
\end{equation}
with $\mu_N$ the nuclear magneton, and $g_l$ and $g_s$ the orbital and the spin
gyromagnetic factors, respectively.

 The gyromagnetic factors $g(\sigma,I), g_p(\sigma,I)$ and $g_n(\sigma,I)$ are
defined by
\begin{equation}
g(\sigma,I) = \frac{\mu(\sigma,I) }{\mu_N I} = g_p(\sigma,I) + g_n(\sigma,I),
\end{equation}
with $g_\tau(\sigma,I), \tau= p, n$, given by
\begin{equation}
g_\tau(\sigma,I) = \frac{1}{[I(I+1)(2I+1)]^{1/2}} \left(
     g^{\tau}_l <\sigma, I||J^\tau ||\sigma, I> +
     (g^{\tau}_s - g^{\tau}_l) <\sigma, I||S^\tau ||\sigma, I> \right).
\label{g-form}
\end{equation}
In the calculations we use for  $g_l$   the free values
and for $g_s$ the free values damped by the usual 0.75 factor
\cite{BM69}
\begin{equation}
g_l^p = 1  \;\;\;\;\;\;\; g_l^n = 0  \;\;\;\;\;\;\;
g_s^p = 5.586 \times 0.75 \;\;\;\;\;\;\; g_s^n = -3.826 \times 0.75 .
\end{equation}
We emphasize that, 
unlike many other models, the g-factor is directly computed by using 
the many-body wave function (Eq. (\ref{g-form})).
In particular, there is no need to introduce any core
contribution, which is a model-dependent concept.

In general, for proton alignment the contribution
$g^p_l <\sigma, I||J^p ||\sigma, I> $ is large and positive
and we expect an increasing $g$ factor. For neutron alignment
$g^{n}_s <\sigma, I||S^n ||\sigma, I>$ is negative and we
therefore expect a decreasing $g$ factor.

In Fig.~7 we present the gyromagnetic factor for the six Dy isotopes along
the yrast band, again with 
the full lines representing the results obtained
with the standard deformation and the dotted lines those obtained with the
enlarged one. Experimental data are presented with their error bars.
Our results show a slight increase at low spins, followed by a clear
reduction at the band crossing region and then a recovery. 
The decrease of the total $g$-factor at the band crossing 
confirms the character of the crossing band
as a neutron aligned band. 
The smoothness of $g$ at high angular momentum indicates a
proton alignment.
The prediction of this trend 
is supported by the $^{154}$Dy g-factor
data \cite{Bir93} which extended the measurement to high spins.
For the low spin part, the PSM predictions are also confirmed
by later experiments \cite{g-Er}. 

In a recent experiment, Alfer {\it et al.} measured g-factors
for $^{158,160,162}$Dy at low spins 
\cite{g-Dy}. In contrast to the behavior
of the Er g-factors at low spins \cite{g-Er}, Alfer {\it et al.} 
found a clear drop in $^{158,160}$Dy 
at spin $I = 6\hbar$, which was not predicted
by the PSM calculation \cite{Sun94}. 
The purpose of our present g-factor calculation is to 
find a possible explanation for the Alfer's data 
when enlarged deformation is studied in this paper. 
However, the result is negative: 
the basic features of the theoretical g-factor do not seem to
change much in our new calculations. 
In the calculation with enlarged deformation, 
one sees only a delayed and smaller decrease of the values at the  
band crossing, but nothing essential has changed for the low spins. 

Bengtsson and \AA berg \cite{Ben86} suggested an increase in g-factor 
at low spins due to changes in deformation and pairing. 
Our microscopic calculations in this paper seem to have the same trend.
The admixture from the $i_{13/2}$ single neutrons to the ground band  
is found to be negligibly small around spin $I = 6\hbar$ in our
calculation, thus
can not be the reason causing the g-factor drop, as suggested by the
authors of Ref. \cite{g-Dy} 
(If this argument is right, one would expect a larger drop of
g-factor for the state $I = 8\hbar$).
To our knowledge, there have been no microscopic calculations 
that can reproduce  this variation of the g-factor around spin $6\hbar$. 
A similar situation is found in $^{50}Cr$ \cite{Pinedo}, when
studied using the spherical shell model (complete $fp$ shell) and the HFB
method. Both predictions are nearly  identical, but they do not agree with
the data \cite{Pakou}, noticeably for $I=4\hbar$, where the g-factor drops
in a way similar to that found in Dy isotopes \cite{g-Dy}. 
We notice that recent measurements do not find this g-factor dropping at
low spins in $^{50}Cr$ \cite{Benc99}

\section{Conclusions}

We have presented a study of the yrast band in  the $^{154-164}$Dy 
isotopes
using the Projected Shell Model (PSM). We have shown that the use
of an input deformation $20\%$ larger than the standard value,
coupled with a slight reduction in effective charge,
leads to an improved description of the yrast band energies,
and of $B(E2)$ values and
transition quadrupole moments in these isotopes. 
The dependence of the $B(E2)$ values on the angular momentum is also 
better described when the larger deformations are used.

We have discussed the changes in the distribution 
of single-particle occupation implied by increased deformations for 
the unperturbed rotational bands.
The different Nilsson single particle energies at enlarged deformations
and the associated changes in the Fermi level were shown to be
the main source of changes in the yrast spectra and the wave functions. 
Appropriate modifications of the Nilsson parameters would have had a
similar effect. 

While general rotational features and the physics related to the band
crossings are well described by the present study, 
limitations of the model are seen from the discussions. 
The lighter isotopes $^{154}$Dy and $^{156}$Dy exhibit softness against 
rotation not present in the calculations.  
A correct description of these nuclei 
would require a study with a richer Hilbert space 
than is contained in this simple PSM.
The observed reduction in g-factor around $I= 6\hbar$ can not be 
explained by the PSM. 

After this manuscript was finished, we become aware of a recent extension
of the PSM by Sheihk and Hara \cite{Shei98} that includes
$\gamma-$deformation in the basis states and performs 3-dimensional
angular momentum projection. Although their preliminary code works in a
very limited model space, great improvement for the description of the
moment of inertia at low spins in rare earth nuclei with neutron number
around 90 is obtained.
This could remove the discrepancies found in our present paper for
$^{154}$Dy and $^{156}$Dy and strongly extend the predictive power of the
PSM. 

We finally emphasize that  
a systematic study for a chain of nuclei is always a 
serious test for microscopic models, before they can 
be used to explain or predict an isolated event in specific isotopes.  

\acknowledgements

This work was supported in part by Conacyt (Mexico) and the National
Science Foundation. Yang Sun acknowledge the hospitality of the Instituto
de Ciencias Nucleares, UNAM, where the final version of this article was
completed.

\newpage

\centerline{\bf Figure Captions}

\bigskip

Figure 1: Angular frequency $\omega$ vs. angular momentum 
$I$ for $^{154,156,158,160,162,164}$Dy are shown in inserts (a,b,c,d,e,f)
respectively. The PSM results using standard deformations are presented as
solid lines; those with enlarged deformations as dotted lines.
Experimental data are represented by diamonds. 

Figure 2: Twice the inertia moment $\Theta$ vs. the square of the
angular velocity $\omega$. The same convention as Fig. 1 is used.

Figure 3: Unperturbed rotational bands for $^{158}Dy$ at standard (a)
and enlarged (b) deformation. For the 2-qp bands the notation
[N lj m] is used for each qp component. 4qp means a four quasiparticle
state.

Figure 4: Nilsson neutron single particle energies around the Fermi level
(represented with a diamond) for standard (a) and enlarged (b)
deformation.

Figure 5: B(E2) values, in $e^2 b^2$, as a function of the angular
momentum $I$. The same convention as Fig. 1 is used.

Figure 6: Reduced transition quadrupole moments 
$Q_t (I\rightarrow I-2)$ 
normalized with respect to $Q_t (2\rightarrow 0)$. 
The same convention as Fig. 1 is used.

Figure 7: G-factors vs. angular momentum I.
The same convention as Fig. 1 is used.

\begin{table}
\begin{center}
\caption{Standard ($\epsilon$) and enlarged ($\epsilon '$) deformations
used in this work for the six Dy isotopes listed in the first column.}
\vspace{0.5cm}
\begin{tabular} {|c|c|c|}
\hline
Isotope & $\epsilon_2$ & $\epsilon_2'$  \\ \hline
$^{154}Dy$ & 0.192 & 0.250  \\
$^{156}Dy$ & 0.217 & 0.270  \\
$^{158}Dy$ & 0.242 & 0.286  \\
$^{160}Dy$ & 0.250 & 0.290  \\
$^{162}Dy$ & 0.258 & 0.308  \\
$^{164}Dy$ & 0.267 & 0.280  \\
\hline
\end{tabular}
\end{center}
\end{table}

\end{document}